\documentclass[aps,prl,superscriptaddress,amsmath,amssymb,a4paper,twocolumn,longbibliography]{revtex4-1}

\usepackage{graphicx}
\usepackage{color}

\usepackage{comment}
\usepackage{booktabs}

\begin{document}

\title{Buckling of self-assembled colloidal structures}

\author{Simon Stuij}
\affiliation{Institute of Physics, University of Amsterdam, Science Park 904, 1098 XH Amsterdam, the Netherlands}
\author{Jan Maarten van Doorn}
\affiliation{Physical Chemistry and Soft Matter, Wageningen University, Stippeneng 4, 6708 WE Wageningen, The Netherlands}
\author{Thomas Kodger}
\affiliation{Physical Chemistry and Soft Matter, Wageningen University, Stippeneng 4, 6708 WE Wageningen, The Netherlands}
\author{Joris Sprakel}
\affiliation{Physical Chemistry and Soft Matter, Wageningen University, Stippeneng 4, 6708 WE Wageningen, The Netherlands}
\author{Corentin Coulais}
\affiliation{Institute of Physics, University of Amsterdam, Science Park 904, 1098 XH Amsterdam, the Netherlands}
\author{Peter Schall}
\affiliation{Institute of Physics, University of Amsterdam, Science Park 904, 1098 XH Amsterdam, the Netherlands}

\begin{abstract}
Although buckling is a prime route to achieve functionalization and synthesis of single colloids, buckling of colloidal structures---made up of multiple colloids---remains poorly studied. Here, we investigate the buckling of the simplest form of a colloidal structure, a colloidal chain that is self-assembled through critical Casimir forces. We demonstrate that the mechanical instability of such a chain is strikingly reminiscent of that of classical Euler buckling but with thermal fluctuations and plastic effects playing a significant role. Namely, we find that fluctuations tend to diverge close to the onset of buckling and that plasticity controls the buckling dynamics at large deformations. Our work provides insight into the effect of geometrical, thermal and plastic interactions on the nonlinear mechanics of self-assembled structures, of relevance for the rheology of complex and living matter and the rational design of colloidal architectures.
\end{abstract}
\maketitle

\emph{Introduction.---} Due to recent advances in colloidal synthesis and interaction control, colloidal self-assembly has become a promising platform for designer materials with controlled internal architecture and tunable physical properties \cite{Manoharan2015,Morphew2017,Gong2017}, such as unprecedented photonic \cite{Lopez2011}, shape-changing \cite{Shah2014,Yan2016} and mechanical properties~\cite{Suzuki2016}. Self-assembled colloidal structures also form excellent model systems to describe complex and biological materials like gels \cite{Zaccarelli2007,VanDoorn2018}, biological cell membranes \cite{Wel2016} and filaments \cite{Li2010,Vutukuri2012}, or flocking behavior \cite{Palacci2013}.
To date, there has been an extensive focus on the dynamical and structural aspects of self-assembly \cite{Meng2010, Zeravcic2014}, but the effective mechanical properties, and in particular mechanical instabilities of self-assembled objects are largely unexplored. Yet, such instabilities play an important role in the response of soft materials, from biological networks~\cite{Broedersz2014} to mechanical metamaterials~\cite{Bertoldi2017}. Semi-flexible biofilaments, polymers and biological shells have been shown to undergo signatures of mechanical instabilities \cite{Dogterom1997}, on which thermal excitations can have an important effect \cite{Paulose2012,Mao2015,Baczynski2008,Pilyugina2017}.
However, a comprehensive understanding of these instabilities in synthetic architectures such as colloidal assemblies is still lacking. In particular, potentially crucial factors such as the effective elastic interactions, the role of geometric non-linearities, stochastic noise and plasticity are virtually unexplored.

\begin{figure}[b!]
    \includegraphics[width=\columnwidth]{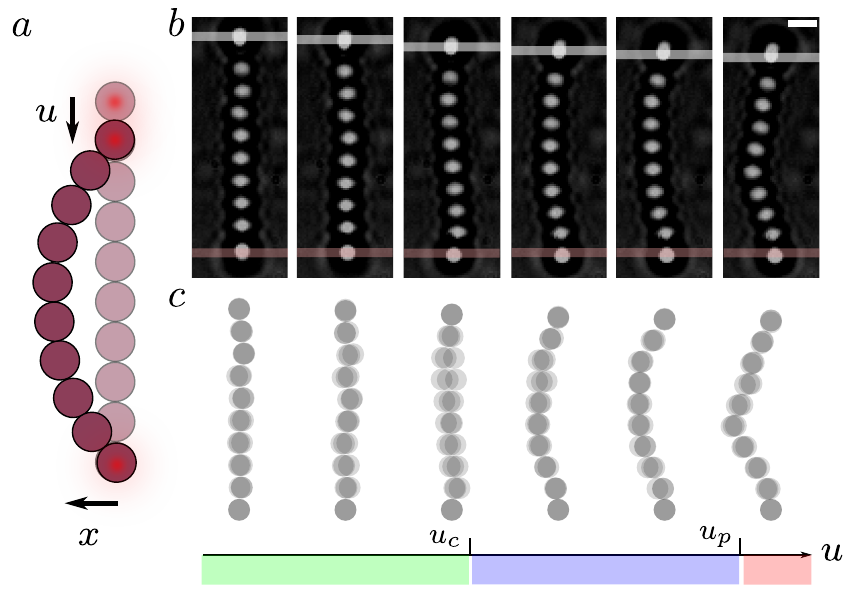}
    \caption{Buckling of a colloidal chain. (a) Sketch of the colloidal chain compressed by optical tweezers (red dots). (b) Bright-field microscope images of the chain at $u=-0.45, -0.25, 0.2, 0.6, 0.8, 1.0 \mathrm{\mu m}$, respectively from left to right, under a compressive displacement. White bars indicate the position of the laser trap that is slowly displaced and red bars the position of the static trap. The scale bar is $3 \mathrm{\mu m}$. The color code demarcates regions of the straight, elastically and plastically buckled chain, bounded by $u_c$ and $u_p$, respectively. (c) Overlay of three reconstructed chains, one corresponding to the still shown in panel (b), one taken 1.5 seconds earlier (light grey) and one 1.5 seconds later (dark grey).}
    \label{fig1}
\end{figure}

Here, we focus on the simplest and most widespread form of a mechanical instability on the simplest form of a self-assembled structure: the buckling of an initially straight colloidal chain upon a compressive load. Combining optical tweezer and microscopy experiments, molecular dynamics simulations and theory, we observe that such chain undergoes a well-defined elastic buckling instability upon compression, close to which thermal bending fluctuations tend to diverge. We further observe critical slowing down: the time scale of the fluctuations diverges at buckling. Molecular dynamics simulations reproduce this behavior quantitatively and allow identifying the critical exponents as the mean field exponents. Finally, we show analytically that a simple continuum model exhibits an analogous divergence of fluctuations, demonstrating the generality of the observed phenomenon.
These results, uncovering the nature of mechanical instabilities in self-assembled structures, provide a crucial step towards understanding the complex mechanics of soft architectures, central to the mechanical function of biological materials and the design of functional colloidal materials.

\emph{Experimental protocol.---} Our system consists of copolymer particles~\cite{Kodger2015} that we assemble into chains using temperature-dependent critical Casimir attractions~\cite{Stuij2017}. The attractive force arises from the confinement of fluctuations of a binary solvent between the surfaces of the colloidal particles. We use particles with a radius of $r=1.25 \mu \mathrm{m}$ suspended in a binary solvent of lutidine and heavy water with lutidine weight fraction $c_L = 0.32$, in which they sediment into a quasi two-dimensional layer. Salt ($5 \mathrm{mM}$ potassium chloride) is added to screen the electrostatic repulsion. By setting the temperature to $\Delta T = 5.5^\circ \mathrm{C}$ below the critical temperature $T_c = 33.6^\circ \mathrm{C}$, we induce an attraction with potential depth $E \sim 10 k_B T$ and range $\sim 0.01 r$ that causes assembly of the particles. We select assembled colloidal chains and use optical tweezers to grab their ends. These chains are initially never straight; to create a straight chain, we lower the temperature to $\Delta T = 6.5^\circ \mathrm{C}$ below $T_c$ to decrease the strength of the critical Casimir interaction, stretch the chain by increasing the separation of the optical tweezers by $\sim 0.5 \mathrm{\mu m}$ and, after straightening, increase the temperature back to $\Delta T = 5.5^\circ \mathrm{C}$. We then apply a compressive displacement $u$ by moving one of the optical tweezers at a constant rate of $27 \mathrm{nm/s}$. We image the individual particles at a frame rate of $20 s^{-1}$, and locate their centers in the image plane with an accuracy of $20 $nm using particle-tracking software \cite{allan_2016_60550}. In addition, we measure the force exerted on the chain from the bead displacement out of the static trap using $F=k(y-y_{trap})$, where $k=0.9 \pm 0.1 \mathrm{pN/\mu m}$ is the trap stiffness (see SI for calibration \cite{supmat}), and $y$ and $y_{trap}$ are the positions of the trapped bead and trap center, respectively. We define $L$ as the end-to-end distance of the chain, and $L_0=24.7 \pm 0.1 \mathrm{\mu m}$ as the end-to-end distance for vanishing force $F=0$.

\emph{Euler buckling.---}
To investigate its buckling behavior, we subject the initially straight colloidal chain to continuously increasing compression. We observe that the chain undergoes a sharp buckling transition at a well-defined compressive displacement $u_c$, as shown in Fig. \ref{fig1}b. In the vicinity of $u_c$, fluctuations significantly increase, as has been predicted theoretically \cite{Bedi2015}. The increasing fluctuations are clearly visible in the superposition of three reconstructed images in Fig. \ref{fig1}c. After buckling, upon further compression, the fluctuations decrease again, and finally, a kink appears at a well-defined large compressive displacement $u_p$.

\begin{figure}[t!]
    \includegraphics[width=\columnwidth]{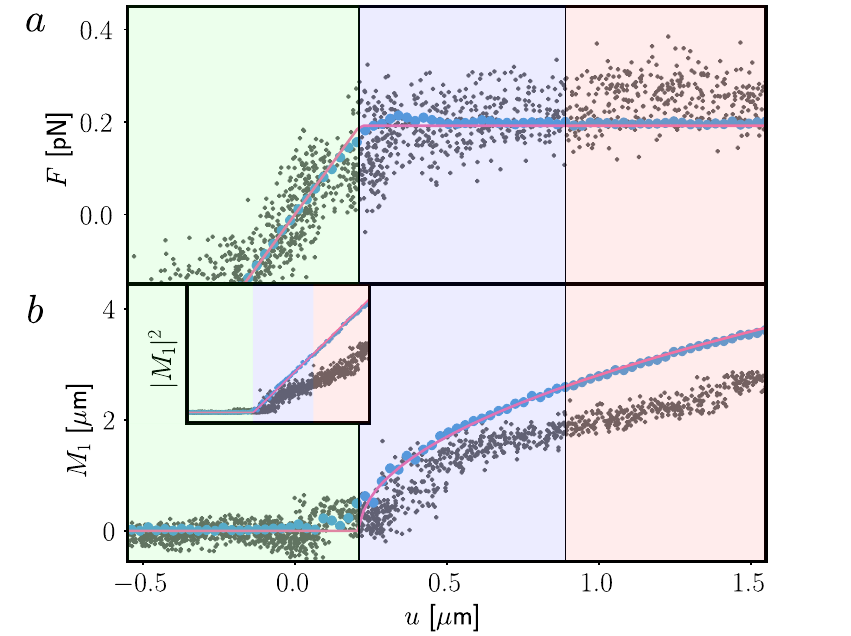}
   \caption
    {Elastic buckling regime: bending force and first Fourier mode, experiments (grey dots) and simulations (blue dots) and continuum model (purple line) (a) Compressive force $F$ exerted by the tweezer on the chain versus displacement, $u$. Note that all experimental data points are depicted, while the simulation data has been averaged over fixed $u$.
    (b) Amplitude of the first Fourier mode $M_1$ of the particle deflections. Only positive mode amplitudes corresponding to positive deflection of the chain are shown. The inset shows the same quantity squared.
    }
    \label{fig2}
\end{figure}

To further elucidate this buckling behavior, we measure the force exerted by the trap on the chain as a function of the compressive displacement $u$ (Fig.~\ref{fig2}a). We observe a linear increase up to a critical displacement $u_c$ beyond which the force remains essentially constant. Such force-displacement curve is strongly reminiscent of a classical Euler buckling problem \cite{Hutchinson1970,Bazant2009,Coulais2015}. To confirm the validity of this analogy, we map our result onto that of a continuous beam.
We use the Euler buckling criterion for the critical force $F_c=\pi^2 B/L_0^2$ and the critical displacement $u_c=F_c / S$, where $B$ is the bending modulus and $S$ the linear stiffness of the beam. Determining the critical force $F_c=0.19 \pm 0.02 \mathrm{pN}$ and displacement $u_c=0.21 \pm 0.02\mu\mathrm{m}$ by interpolation, we find that the bending rigidity of the chain is $B= 11.9\pm 1 \mathrm{pN \mu m^2}$ and the the linear stiffness is $S=0.9 \pm 0.1 \mathrm{pN/\mu m}$. Furthermore, the value of the stiffness $S$ is consistent with that obtained from a linear fit to the pre-buckling slope. Such an excellent agreement between a model for athermal slender structures and our thermally activated colloidal chain is striking.

The validity of this mapping is further confirmed by the shape of the buckled state, which we quantify by the amplitude $M_1$ of the first Fourier mode of the beam deflection (see SI \cite{supmat}) as a function of the compressive displacement $u$ (Fig.~\ref{fig2}b).
While this amplitude is close to zero in the pre-buckling regime, $u<u_c$, it sharply departs from zero and increases as $M_1 \propto (u-u_c)^{1/2}$ beyond the buckling point, see Fig. 2b inset. Again, this result is qualitatively similar to that of a macroscopic Euler buckling problem~\cite{Hutchinson1970,Bazant2009,Coulais2015}.
Note that such deflection-displacement curve provides an independent measurement of the critical displacement $u_c=0.21\pm 0.02 \mu\mathrm{m}$, which is equal to the previous measurement within experimental errors.
These results are consistent with and rationalize previous studies reporting a bending rigidity of linear assembled structures~\cite{Dinsmore2006,Pantina2005,Biswal2003}. Interestingly, the first Fourier amplitude starts to deviate from zero already before the buckling transition (Fig.~\ref{fig2}b). As we will see in the following, this is related to the increasing fluctuations of the chain approaching the buckling transition.

\begin{figure}[t!]
    \includegraphics[width=\columnwidth]{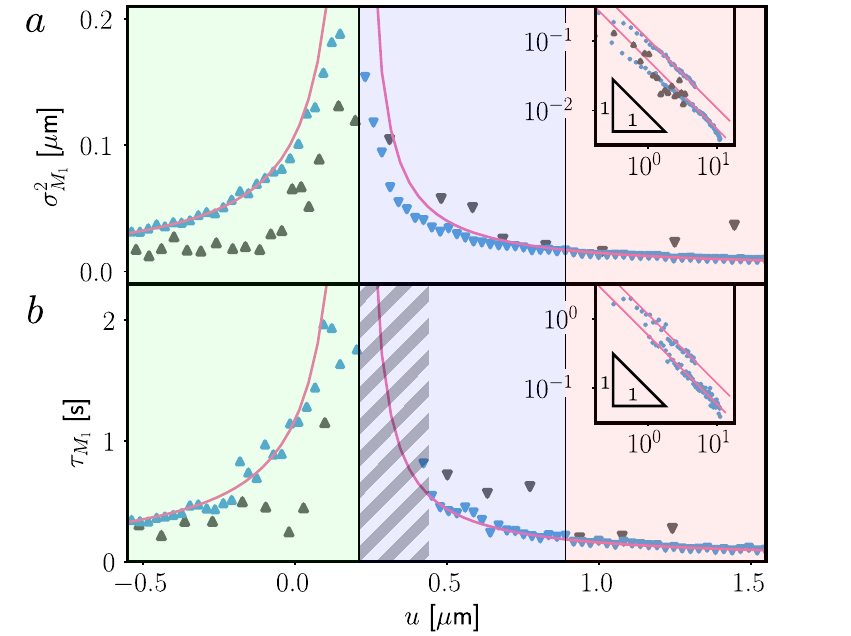}
   \caption
    {Fluctuations close to buckling.
    (a) Variance $\sigma^2$ of $M_1$ above the mode of its distribution versus the compressive displacement $u$. The inset shows a loglog plot of $\sigma^2$ versus $|u-u_c|/u$.
    (b) Correlation time $\tau$ of $M_1$.
    The shaded area directly after $u_c$ indicates the region where there are two decay times difficult to distinguish. The inset shows a loglog plot of $\tau$ versus $|u-u_c|/u$. Experimental (numerical, continuum model) data is represented by grey triangles (blue triangles, purple lines). Simulation and continuum model values are divided by a factor of $3$ to fit on the same axis.}
    \label{fig3}
\end{figure}

\emph{Fluctuations.---} To analyze these fluctuations in detail, we measure the variance $\sigma_{M_1}$ of the first Fourier amplitude $M_1$ around its mean value, defined by $\sigma^2_{M_1}=\langle(M_1-|\bar{M_1}|)^2\rangle_{M_1 \geq |\bar{M_1}|}$. Upon approaching the buckling point, this variance grows and diverges (Fig.~\ref{fig3}a).
The double-logarithmic plot (inset) suggests a divergence $\sigma^2_{M_1} \sim |u-u_c|^{-\nu}$ with exponent $\nu=1$. We also measure the typical time scale of fluctuations, $\tau_{M_1}$, from exponential fits to the decay of the autocorrelation function $C(\Delta t)=\langle M_1(t) M_1(t+\Delta t)\rangle$; this fluctuation time shows likewise a significant increase upon approaching $u_c$, see Fig.~\ref{fig3}b. The uncertainty and limited number of data points do not allow us to pinpoint the divergence of these growing fluctuations quantitatively. Interestingly, higher-order Fourier mode variances do not exhibit any growth upon approaching $u_c$ (data not shown).

\begin{figure}[t!]
\includegraphics[width=\columnwidth]{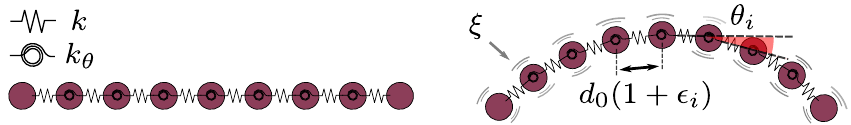}
     \caption{
     Schematic of the model system used for the simulations and the analytical model.
     }
     \label{fig4}
 \end{figure}

\emph{Numerical simulations protocol.---}To rationalize the experimental findings and have access to more precise and better statistics, we perform molecular dynamic simulations of elastically coupled particles in two dimensions subjected to thermal fluctuations, see Fig.~\ref{fig4}. Specifically, we solve the overdamped Langevin equation \cite{Ermak1978}:
\begin{equation}\label{eqn:MDeom}
\dot{\boldsymbol{r}_i}  = -\frac{D}{k_BT}\nabla_{\mathbf{r}_i}V+\sqrt{2D}\xi,
\end{equation}
where $\xi$ a normalized stochastic thermal force, $D=0.138 \pm 0.1 \mathrm{\mu m^2/s}$ is the diffusion coefficient measured experimentally by tracking diffusing colloids, and $T$ the temperature equal to the experimental temperature.  The potential energy is given by:
\begin{equation}\label{eqn:MDpotential}
V = \frac{k}{2} d_0^2 \sum_{i=1}^{N-1}  \epsilon_i^2 +
       \frac{k_{\theta}}{2} \sum_{i=1}^{N-2} (\theta_i-\theta_{i,0})^2,
\end{equation}
with $\epsilon_i$ the extension of bond $i$, $\theta_i$ the angle between bonds $i$ and $i+1$, and $\theta_{i,0}$ the equilibrium angles which are equal to zero for an initially straight chain. The equilibrium bond distance is determined from experiments as the average distance between particles $d_0=L_0/N-1$. We also take the bending rigidity and bond stiffness from the experimental measurements $k=S(N-1)$ and $k_{\theta}=B/d_0$ and assume an infinite trapping potential. We then apply compression by moving the traps stepwise towards each other with a displacement $u_{step}=0.01d_0$ and waiting time $t_{step}=32 d_0^2/D$ between each step. This gives an average compression rate of $0.9 \mathrm{nm/min}$, much slower than the experiments, allowing us to acquire very good simulation statistics at each position.

\emph{Numerical simulations results.---}Despite the simple assumptions of the numerical model, the results are in strikingly good agreement with the experiments (Figs.~\ref{fig2} and~\ref{fig3}). The force, deflection, fluctuation and correlation time vs. displacement curves all predict the buckling instability at $u_c$ and correctly describe the force and fluctuations behavior, lending credence to the simplifying assumptions of the model. The quantitative deviations are likely due to the fact that (i) the experimental boundary conditions (laser traps) do not allow complete free rotations of the trapped colloids and (ii) the real colloidal chain has also nonlinear terms in the bending stiffness.
Nevertheless, the numerical model confirms the growth of fluctuations close to buckling and its better statistics indicates that these fluctuations indeed diverge as $\sigma^2_{M_1} \sim |u-u_c|^{-1}$, see inset of Fig.~\ref{fig3}a. For the correlation times, the simulations find a similar divergence $\tau_{M_1} \sim |u-u_c|^{-1}$, see inset of Fig.~\ref{fig3}b.

\emph{Continuum model.---}
To obtain insight into the critical behaviour of this stochastic buckling transition, we consider a simple---analytically solvable--- continuum limit of Eq. (2), known as the extensible elastica \cite{Magnusson2001}. 
In this limit, the energy can be decomposed into independent contributions from each fourier mode. To first order in $u$, the energy dependence on the first mode amplitude $M_1$ becomes a double-well, given by
\begin{equation}\label{eqn:Econt}
V_1 = \frac{S\pi^2}{4L_0}(u_c-u)M_1^2+\frac{S\pi^4}{32L_0^2} M_1^4+\mathcal{O}(u^2)+\mathcal{O}(M_1^6),
\end{equation}
where $u_c=B\pi^2/SL_0^2$, $B$ the bending rigidity and $S$ the stretching stiffness. Higher modes exhibit a single harmonic energy dependence and equilibrate to zero (see SI for a detailed derivation \cite{supmat}). Mechanical equilibria of this extensible elastica, prescribed by the condition $\partial V_1 / \partial M_1 = 0$, are given by $M_{1,m}=0$ in the pre-buckling regime ($u<u_c$), and by $M_{1,m}=\pm 2/\pi\sqrt{L_0(u-u_c)}$ in the post-buckling regime ($u>u_c$). The corresponding forces, are $F_m=k u/2$ for $u<u_c$ and $F_m=F_c(1+(u-u_c)/2L_0)$ for $u>u_c$, (see SI \cite{supmat}). Furthermore, if we assume that in equilibrium, the bending energies given by Eq.~\eqref{eqn:Econt} obey a Boltzmann distribution, then the mode fluctuations around the average become Gaussian distributed with variance
\begin{equation}\label{eqn:Sigma_ana}
\sigma^2_{M_1}=
\begin{cases}
\frac{2kT L_0}{\pi^2 S}|u_c-u|^{-1} & u < u_c  \\
\frac{kT L_0}{\pi^2 S}|u_c-u|^{-1} & u > u_c.
\end{cases},
\end{equation}
Note that this approach breaks down for $u >\sim u_c$, in the post-buckling regime near the buckling point, where the distribution becomes bimodal rather than a single Gaussian as predicted by Eq.~\ref{eqn:Sigma_ana}. For the fluctuation time, the overdamped dynamics for a square-well predicts that $\tau_{M_1}=\sigma^2_{M_1}/D_{M_1}$, where $D_{M_1}=2D/(N-1)$ is the effective mode diffusion.
These predictions of the scaling are in perfect agreement with the experiments and simulations as shown in Figs.~\ref{fig2} and~\ref{fig3}. Also, the factor 2 difference between the pre- and post-buckling regime is consistent with the numerical results. A physically appealing picture emerges from these results: once in presence of stochastic noise, the classical buckling transition remains a supercritical bifurcation, but the vicinity of the bifurcation is associated with fluctuations of diverging magnitude and timescales.

\begin{figure}
     \includegraphics[width=\columnwidth]{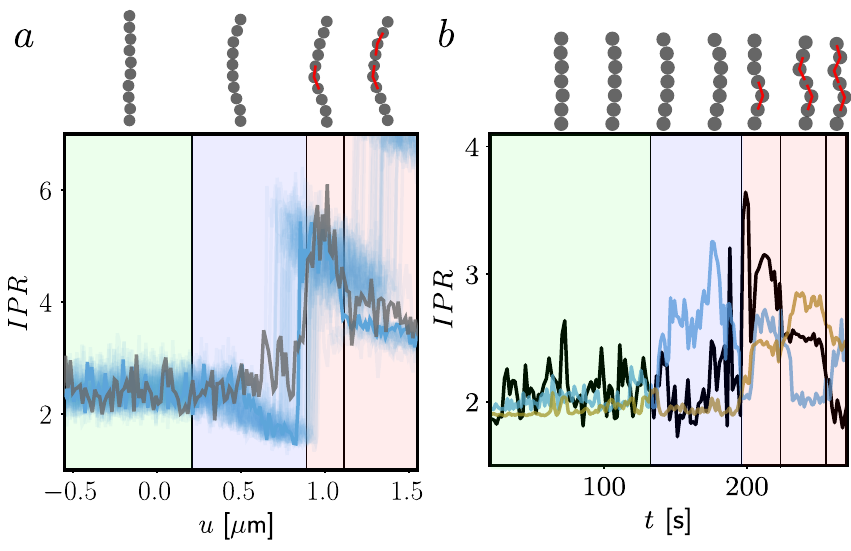}
     \caption{Plastic buckling.
     (a) Inverse participation ratio (IPR) of the experimental chain (grey), and of $50$ independent MD simulation runs (blue shading) as a function of continuously increasing compression.  The simulations are performed with elastic parameters as in Fig.~\ref{fig2} and $\theta_c=0.21~\mathrm{rad}$. Vertical lines and colors distinguish regimes of the straight, elastically and plastically buckled chain, and indicate two plastic slippage events. Reconstructed snapshots show the experimental chain for $u=0$, $0.5$, $1$ and $1.5$ $\mathrm{\mu m}$, with snapped bonds highlighted in red. (b) The IPR (black), $M_1$ (blue) and  $M_2$ (yellow) versus time of a different compression experiment. Here, the chain was shorter ($N=7$), and the trap was moved stepwise, by $\delta u=0.1 \mathrm{ \mu m}$ every $60$s.
     }
     \label{fig5}
\end{figure}

\emph{Plastic buckling.---} At even larger displacements $u>u_p$, the chain undergoes localized bending deformations as shown in Fig. 1b and c (utmost right images), which we find to be irreversible upon releasing the applied compression. To quantify this degree of localization, similar to plastic events in amorphous materials, we calculate the inverse participation ratio (IPR) which varies between $N-2$ for fully localized deformations and $1$ for distributed deformations, as defined by
\begin{equation}\label{eqn:IPR}
IPR = (N-2)\frac{\sum_{i=1}^{N-2}\hat{\theta}^4_i}{(\sum_{i=1}^{N-2}\hat{\theta}^2_i)^2}.
\end{equation}
Here $\hat{\theta}_i=|\theta_i|-\langle|\theta_i|\rangle_{u<u_c}$, i.e. the local
angular deviation from the straight chain. When the chain buckles elastically, the IPR remains small, see grey curve in Fig.~\ref{fig5}a, while at larger compression $u=u_p=0.89 \pm 0.02 \mathrm{\mu m}$, when the chain develops a kink, a clear spike is observed. The value of about 6, which is only slightly smaller than the maximum $N-2=8$ indeed suggests very localized deformations. These features can be easily reproduced in the simulations, when we augment our numerical model with a simple elasto-plastic model. Beyond a threshold angle $\theta_p$, an instant plastic relaxation occurs such that the equilibrium bond angle becomes $\theta_{0,i}=\theta_c$. Taking a value $\theta_p=0.21 \mathrm{rad}$ gives results qualitatively and quantitatively similar to the experiment, see blue shading in Fig.~\ref{fig5}a. By repeating $50$ simulations we obtain an average $u_{p,sim}=0.84 \pm 0.09 \mathrm{mu m}$, which indeed corresponds to the value $u_p$ observed in the experiments; the large variation between different simulation runs shows that also the plastic event is influenced by stochastic noise. Intriguingly, this combination of elasto-plastic dynamics and thermal noise can further lead to higher-order buckling modes when the chain is compressed at higher compression rates (Fig~\ref{fig5}b). We observe a sequence of buckling transitions through mode $1$, mode $2$ and mode $3$, that we interpret as a sequence of plastic events, as clearly shown by the mode 1 and 2 amplitudes (blue and olive) and IPR (black).

\emph{Outlook.---} We have unveiled the rich stochastic buckling dynamics of a colloidal chain under uniaxial compression by combining experiments, simulations and analytic modelling. Remarkably, in the elastic regime, we find that bending fluctuations diverge upon approaching the buckling point. This divergence of fluctuations is described based on elastic bending interactions and stochastic noise. These results have important consequences for the mechanics of soft architectures that are comprised of thermal strands, such as colloidal networks. While the presence of linear elastic response can be understood from the Casimir interactions~\cite{Stuij2017}, the presence of bending interactions and plasticity is surprising and difficult to interpret. We speculate that these stem from intricate contact mechanics~\cite{Pantina2005}, such as finite surface roughness, rolling or sliding frictional effects and charge disparity.
The observed divergence of fluctuations then translates into a maximum entropic contribution to the stress, which could manifest in the rheology of these larger networks. Our results open up unique avenues for self-assembled colloidal structures with advanced nonlinear mechanics of relevance for the understanding of the rheology of gels~\cite{VanDoorn2018}, the mechanics of living tissues ~\cite{Broedersz2014} and of designer colloidal architectures~\cite{Bertoldi2017}.

\emph{Acknowledgements.---} C.C, T.K., J.S. and P.S. acknowledge support by, respectively, Veni, Veni, Vidi and Vici fellowships from the Netherlands Organization for Scientific Research (NWO). J.M.D. acknowledges funding by the Industrial Partnership Program "Hybrid Soft Materials" of Unilever and NWO.

 \bibliographystyle{apsrev4-1}
 \bibliography{StochasticBucklingChain}

\end{document}


\title{Supplementary information of article\\Buckling of self-assembled colloidal structures}

\author{Simon Stuij}
\affiliation{Institute of Physics, University of Amsterdam, Science Park 904, 1098 XH Amsterdam, the Netherlands}
\author{Jan Maarten van Doorn}
\affiliation{Physical Chemistry and Soft Matter, Wageningen University, Stippeneng 4, 6708 WE Wageningen, The Netherlands}
\author{Thomas Kodger}
\affiliation{Physical Chemistry and Soft Matter, Wageningen University, Stippeneng 4, 6708 WE Wageningen, The Netherlands}
\author{Joris Sprakel}
\affiliation{Physical Chemistry and Soft Matter, Wageningen University, Stippeneng 4, 6708 WE Wageningen, The Netherlands}
\author{Corentin Coulais}
\affiliation{Institute of Physics, University of Amsterdam, Science Park 904, 1098 XH Amsterdam, the Netherlands}
\author{Peter Schall}
\affiliation{Institute of Physics, University of Amsterdam, Science Park 904, 1098 XH Amsterdam, the Netherlands}

\maketitle

\section{Optical tweezers}
\begin{figure}[t!]
    \includegraphics[width=0.5\columnwidth]{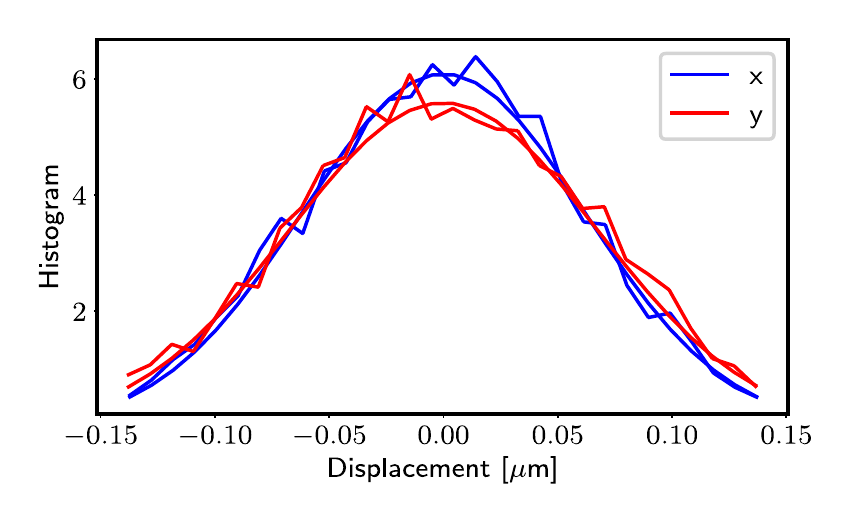}
   \caption
    {Trap calibration. Displacement histogram of a single trapped particle in the x- and y-direction (see legend) imaged at $20$ fps for $6$ minutes resulting in a total of $7200$ images. The best gaussian fit gives standard deviations of $\sigma_x = 0.062 \mathrm{\mu m} $ and $\sigma_y = 0.068 \mathrm{\mu m} $.
    }
    \label{fig1}
\end{figure}

For the optical tweezers, laser light of $1064$nm was used at a power of $20 \pm 5 \mathrm{mW}$. The trap constants were determined by tracking the brownian movement of a single colloidal particle in the trap and fitting its displacements from the trap center with a gaussian distribution to obtain the standard deviation $\sigma_{trap}$, see Fig.~\ref{fig1}.
We used a long measurement time such that the out-of-trap displacements become Boltzmann distributed. Assuming a harmonic trap, the trap constant is then determined by $k_{trap}=kT/\sigma^2_{trap}$. We obtain $k_x = 1.1 \pm 0.2 \mathrm{pN/\mu m}$ and $k_y = 0.9 \pm 0.2 \mathrm{pN/\mu m}$. The error is estimated based on the locating accuracy of $\epsilon_{track} = 0.02 \mathrm{\mu m}$ and calculating the resulting error in determintation of $\sigma_{trap}$ by $\epsilon^{\sigma}_{trap}=\epsilon^2_{track}/\sigma_{trap}$.
The partial absorbtion of the laser light by the binary solvent causes a local heating of $~0.5 \mathrm{K}$ at the trap. This was determined by measuring the temperature at which phase separation occurs in the laser focus and subtracting that from the phase separation temperature when the laser is turned off. This temperature increase is expected to cause a slight increase of the critical Casimir attraction close to the trapped bead.

\section{Mode definition and effective mode diffusion constant}
After locating the particles in the chain a mode decomposition is performed in the following manner: First, the out-of-line deflection of every non-trapped particle $j$ is calculated as the perpendicular distance $x_j$ from the line connecting the two trapped particles at the ends. Next, a scaled discrete sine transform of type 1 is performed on $x_j$ defined by:
\begin{equation}\label{eqn:dst}
M_i =\frac{2}{N-1} \sum_{j=1}^{N-2} x_j \sin(\frac{\pi}{N-1} j i ) \quad i=1,...,N-2
\end{equation}
Here the normalization has been chosen such that:
\begin{equation}\label{eqn:reverse_dst}
x_j = \sum_{i=1}^{N-2} M_i \sin( i j \frac{\pi}{N-1}) \quad j=0,...,N-1
\end{equation}

Based on the overdamped langevin equation of individual colloids, Eq.~(1) in the main text, we can derive an equivalent dynamical equation in terms of modes, given by:
\begin{equation}
\dot{M_i}  = -\frac{D_M}{k_BT}\nabla_{M_i}V+\sqrt{2D_M}\xi.
\end{equation}
Here $D_M$ is an effective diffusion coefficient. This coefficient is equal for all modes and can be derived by inserting Eq.~(\ref{eqn:reverse_dst}) in Eq.~(1) of the main text. This gives $D_M = 2D/(N-1)$.

\section{Molecular dynamics simulations}
Colloidal particles with position $\boldsymbol{r_i}$ in an assembled chain satisfy the Langevin equation:
\begin{equation}
m\ddot{\boldsymbol{r}_i}  = -\frac{kT}{D} \dot{\boldsymbol{r}}-\nabla_{\mathbf{r}_i}V+\sqrt{2 D}\xi
\end{equation}
with $D=kT/\gamma$ the diffusion coefficient, $\gamma$ the viscous drag coefficient, $\xi$ a normalised stochastic force, and
\begin{equation}\label{eqn:MDpotential}
V = \frac{k}{2} d_0^2 \sum_{i=1}^{N-1}  (\epsilon_i-1)^2 +
       \frac{k_{\theta}}{2} \sum_{i=1}^{N-2} (\theta_i-\theta_{i,0})^2,
\end{equation}
where $d_0$ is the equilibrium bond distance, $\theta_{i,0}$ the equilibrium angle,  $\epsilon_i = |\boldsymbol{r}_{i+1}-\boldsymbol{r}_{i}|/d_0-1$ and $\cos(\theta_i) = (\boldsymbol{r}_{i+2}-\boldsymbol{r}_{i+1}) \cdot(\boldsymbol{r}_{i+1}-\boldsymbol{r}_{i})/|\boldsymbol{r}_{i+2}-\boldsymbol{r}_{i+1}||\boldsymbol{r}_{i+1}-\boldsymbol{r}_{i}|$.
On timescales $\Delta t>\frac{mD}{kT}$ the Langevin equation can be considered overdamped and reduces to
\begin{equation}\label{eqn:MDeom}
\dot{\boldsymbol{r}_i}  = -\frac{D}{k_BT}\nabla_{\mathbf{r}_i}V+\sqrt{2D}\xi.
\end{equation}
These can be simulated by molecular dynamics simulations following the Ermak-McCammon equation [31]:
 \begin{equation}
\boldsymbol{r}_i(t+\Delta t)-\boldsymbol{r}_i(t) = -\frac{D}{kT}\nabla_{\mathbf{r}_i}V \Delta t + \sqrt{2 D \Delta t}\xi,
 \end{equation}
We simulate an infinitely stiff trap by fixing the positions of the two end particles.
Trap movement is then implemented by moving one end particle towards the other.

Time, length and energy were expressed in natural units such that $t_D=d_0^2/D=1 \mathrm{t_D}$, $kT=1 \mathrm{kT}$, $d_0=1 \mathrm{d_0}$.
In all simulations the timestep was set to $\Delta t=2^{-16}\mathrm{t_D}$ which is small enough to have a stable integration.
In order to compare with experiments all quantities where later rescaled using the experimental values $D=0.138 \mathrm{\mu m^2/s}$, $kT= 4.14 \times 10^{-21} \mathrm{J}$ corresponding to $\sim 0.004 \mathrm{pN \mu m}$, and $d_0=2.74 \mathrm{\mu m}$.

\subsection{Simulation parameters and protocol}
The experimental values of the bending and stretching stiffness as determined from the experimental force are, respectively, $k_{\theta}=1048 \mathrm{kT}$, and $k=14760 \mathrm{kT/d_0^2}$. To simulate the compression experiment, we moved the trap in 128 steps of $\Delta u=0.01 \mathrm{d_0}$ starting from $u_i=-0.3 \mathrm{d_0}$, with a waiting time of $t_{step}=32\mathrm{t_D}$ at each step. This gives a similar total displacement as the experiment. The compression speed is much lower than in the experiments in order to obtain better statistics. For the analysis, we disregarded the first $8\mathrm{t_D}$ after each trap displacement to allow for equilibration. In experimental units, this entire ramp translates to a total displacement of $u_{tot}=3.5 \mathrm{\mu m}$ over a time of $62~ \mathrm{hours}$ with each step taking $30~\mathrm{min}$, giving an effective speed of $v_{trap}=0.9\mathrm{nm/min}$.

\subsection{Elasto-plastic simulations}
In the elasto-plastic simulations, we allowed an instant plastic relaxation to occur at a threshold angle $\theta_p$, such that the new equilibrium bond angle becomes $\theta_{0,i}=\theta_c$. For these simulations, we used the same $k_{\theta}$ and $k$ value as for the previous elastic simulations. A continuously increasing trap displacement was simulated for a total time of $t_{tot}=1.9 \mathrm{t_D}$, increasing from $u_{i} = -0.3 \mathrm{d_0}$ to $u_f = 2.5 \mathrm{d_0}$. This translates to a trap speed of $v_{trap}=37 \mathrm{nm/s}$, close to the actual experimental value. In order to estimate the critical bending angle, we performed $50$ simulation runs at a number of $\theta_c$ ranging from $0.1$ to $0.2\mathrm{rad}$, with steps of $0.01 rad$. The best fitting $\theta_c$ was determined by comparing the average plastic compression $u_p$ to the experimental value. This gave $\theta_c=0.21 rad$.

\section{Theoretical model details}

\subsection{Deriving the double well from the extensible elastica}
We parameterize the shape of an extensible elastic by $\boldsymbol{r}(s)=(x(s),y(s))$ with $s$ running from $0$ to $L_0$, the rest length. The compressive strain is defined as $\gamma(s)=\sqrt{(dx/ds)^2+(dy/ds)^2}$ and the orientation angle as $\phi(s)=\arctan(d y/d x)$.
The energy functional of an elastica including elastic energy and work exerted by a compressive force $F$ is given by [32]
\begin{equation}\label{eq:Eelastica}
V = \frac{B}{2} \int_0^{L_0}  \left(\frac{d\phi}{ds}\right)^2 ds +
        \frac{L_0 S}{2 } \int_0^{L_0} (\gamma-1)^2 ds +  F\left(\int_0^{L_0}\gamma \cos(\phi)ds - R\right).
\end{equation}
Here $R=L_0-u$ is the end-to-end length, $B$ is the bending rigidity and $S$ the stretching stiffness of the elastica.
Minimizing $V$ with respect to $\gamma$ and $F$, we find
\begin{equation}\label{eq:gamma_and_force}
\gamma = 1 - \frac{F}{SL_0} \cos{\phi}, \quad F = SL_0\frac{\int_0^{L_0} \cos{\phi} ds-R}{\int_0^{L_0} (\cos{\phi})^2ds}
\end{equation}
Inserting these back into Eq.~(\ref{eq:Eelastica}), we obtain an energy $V_{\phi}$ purely as function of the orientation angle, given by
\begin{equation*}
V_{\phi} = \frac{B}{2} \int_0^{L_0}  \left(\frac{d\phi}{ds}\right)^2 ds
    +\frac{S L_0}{2}\frac{\int_0^{L_0} \cos({\phi}) ds - R}{\int_0^{L_0} (\cos{\phi})^2 ds}
\end{equation*}
This is the energy we will use to determine the equilibrium angles $\phi(s)$ and also the size of thermal fluctuations in $\phi$.
Note that it is indeed correct to use $V_{\phi}$ to determine the equilibrium.
However, using $V_{\phi}$ to determine the size of fluctuations disregards the effect of thermal fluctuations in $\gamma$ and $F$.
These fluctuations are not uncoupled from fluctuations in $\phi$, as can be seen from Eq.~(\ref{eq:Eelastica}).
Yet, we assume that these fluctuations have negligable influence.

After a Fourier transform assuming Neumann boundary conditions $\phi = \sum_{n=1}^{\infty}\alpha_n \cos(\frac{n\pi}{L_0}s)$, $V_{\phi}$ decomposes into $V_{\phi}=Su^2/2+\sum V_{\alpha_n}$. It follows that up to a critical compression $u_c$ all modes equilibrate to zero. After $u_c$, the first mode becomes nonzero and the chain buckles, which can be seen from
\begin{equation}\label{eq:Valpha1}
V_{\alpha_1}    = \frac{SL_0}{4}\left(u_c^0+\frac{u^2}{L_0}-u\right)\alpha_1^2
     +\frac{S}{32}\left(L_0^2-\frac{7u L_0}{2}-2u^2\right)\alpha_1^4 +\mathcal{O}(\alpha_1^6),
\end{equation}
where $u^0_c=\pi^2 B/SL_0^2$. The buckling compression of the first mode is found by determining the root of the term in front of $\alpha_1^2$, giving $u_c=\frac{L_0}{2}\left(1-\sqrt{1-4u^0_c/L_0}\right)$.
In the regime that we probe experimentally, $u_c/L_0$ and $u/L_0$ are small numbers. Therefore, $u_c \approx u_c^0$, and lowest order terms dominate in $V_{\alpha_1}$, which reduces to
\begin{equation}\label{eq:Valpha1_stiff}
V_{\alpha_1} = \frac{SL_0}{4}\left(u_c^0-u\right)\alpha_1^2+\frac{SL_0^2}{32}\alpha_1^4 +\mathcal{O}(u_c^2)+\mathcal{O}(u^2)+\mathcal{O}(\alpha_1^6)
\end{equation}
Minimizing this energy we see that the equilibrium first mode is given by
\begin{equation}\label{eq:alpha1m}
\alpha^2_{1,m} =
\begin{cases}
0 & u < u_c  \\
\frac{4}{L_0}\Delta u +\mathcal{O}(\Delta u^2) & u > u_c,
\end{cases}
\end{equation}
with $\Delta u = u- u_c$.

To derive the compressive force up to first order in $\Delta u$, care has to be taken to solve $\alpha^2_{1,m}$ from Eq.~(\ref{eq:Valpha1}) one order higher in terms of $u_c/L_0$. Doing that and inserting in Eq.~(\ref{eq:gamma_and_force}) one obtains
\begin{equation}
F_m =
\begin{cases}
S u  & u < u_c  \\
F_c ( 1+\Delta u / 2 L_0 ) +\mathcal{O}(\Delta u^2) & u > u_c.
\end{cases}
\end{equation}
where $F_c=Su_c$.

As a last step to obtain the double-well potential stated in the main text we have to transform $\alpha_1$ to $M_1$, defined by Eq.~(\ref{eqn:dst}). Using that for small deflection and compressions we have $M_1 = L_0 \alpha_1/\pi$, and Eq.(~\ref{eq:Valpha1_stiff}) becomes Eq.(~$3$) of the main text.